\documentclass{article}

     \usepackage[preprint]{neurips_2025}

\usepackage[utf8]{inputenc} 
\usepackage[T1]{fontenc}    
\usepackage{hyperref}       
\usepackage{url}            
\usepackage{booktabs}       
\usepackage{amsfonts}       
\usepackage{nicefrac}       
\usepackage{microtype}      
\usepackage{xcolor}         
\usepackage{graphicx} 
\usepackage{amsmath}  
\usepackage{amssymb}  
\usepackage{algorithm} 
\usepackage{algpseudocode} 
\usepackage{tikz} 
\usetikzlibrary{shapes.geometric, arrows.meta, positioning} 

\definecolor{airforceblue}{rgb}{0.36, 0.54, 0.66}
\usepackage{wrapfig}
\usepackage{titlesec}
\titlespacing{\section}{0pt}{*1}{*0.5}
\titlespacing{\subsection}{0pt}{*0.8}{*0.3}

\title{Understanding Galaxy Morphology Evolution Through Cosmic Time via Redshift Conditioned Diffusion Models}

\author{%
  Andrew Lizarraga$^{1}$,
  Eric Hanchen Jiang$^{1}$,
  Jacob Nowack$^{2}$, \\
  {\bf Yun Qi Li}$^{3}$,
  {\bf Ying Nian Wu}$^{1}$,
  {\bf Bernie Boscoe}$^{2}$,
  {\bf Tuan Do}$^{4}$\\
  \\
  $^{1}$Department of Statistics and Data Science, UCLA\\
  $^{2}$Department of Computer Science, Southern Oregon University\\
  $^{3}$Department of Physics and Astronomy, University of Washington \\
  $^{4}$Department of Physics and Astronomy, UCLA\\ 
  \\
  \texttt{ \{andrewlizarraga, ericjiang0318, yunqil\}@g.ucla.edu},\\
  \texttt{\{boscoeb, nowackj\}@sou.edu },
  \texttt{ywu@stat.ucla.edu}, \texttt{tdo@astro.ucla.edu }\\
  \url{https://github.com/astrodatalab/lizarraga_2024}\\
}

\begin{document}

\maketitle

\begin{abstract}
    Redshift measures the distance to galaxies and underlies our understanding of the origin of the Universe and galaxy evolution. Spectroscopic redshift is the gold-standard method for measuring redshift, but it requires about $1000$ times more telescope time than broad-band imaging. That extra cost limits sky coverage and sample size and puts large spectroscopic surveys out of reach. Photometric redshift methods rely on imaging in multiple color filters and template fitting, yet they ignore the wealth of information carried by galaxy shape and structure. We demonstrate that a diffusion model conditioned on continuous redshift learns this missing joint structure, reproduces known morphology-$z$ correlations.
    We verify on the  HyperSuprime-Cam survey, that the model captures redshift-dependent trends in ellipticity, semi-major axis, Sérsic index, and isophotal area that these generated images correlate closely with true redshifts on test data. To our knowledge this is the first study to establish a direct link between galaxy morphology and redshift. Our approach offers a simple and effective path to redshift estimation from imaging data and will help unlock the full potential of upcoming wide-field surveys.
\end{abstract}

\section{Introduction}
\label{sec::introduction}

Understanding the intricate processes of galaxy formation and evolution represents a frontier challenge in modern astrophysics. Galaxies are not static entities but dynamic systems that undergo profound transformations in morphology, structure, and composition over cosmic times. Fundamental to these studies, redshift (denoted $z$) is an indispensable observational measurement for cosmic time and distance, enabling the study of galaxies at different evolutionary stages. However, the inability to continuously observe individual galaxies throughout their multi-billion-year lifespans imposes a fundamental limitation, restricting empirical studies to statistical ensembles from discrete temporal snapshots, e.g. using redshift. Additionally, unlike photometric surveys such as the Sloan Digital Sky Survey (SDSS)~\cite{ahn2012ninth}, the Dark Energy Survey (DES)~\cite{abbott2021dark}, the Kilo-Degree Survey (KiDS)~\cite{kuijken2019kids}, and the Hyper Suprime-Cam (HSC) Survey~\cite{aihara2019second}, comprised of potentially millions or billions of galaxies, redshift-matched surveys, such as \cite{do2024galaxiesml} often comprise of a few 100,000 samples due to the computational time, cost, and effort of computing the redshift.
To the author's knowldedge there is still no reliable method for predicting redshift given photometric images due to concerns of deep learning models not learning to utilize physical properties of galaxy structure in order to make a prediction.
Here, we investigate the potential of generative models, conditioned on redshift, to bridge this observational chasm by synthesizing high-fidelity galaxy populations across cosmic time, thereby offering a novel computational lens for simulating and understanding their evolutionary trajectories \citep{galaxy_dreams, foundation_models}.

Among emerging generative techniques, Denoising Diffusion Probabilistic Models (DDPMs) have demonstrated exceptional capabilities in producing highly realistic imagery across diverse scientific and visual domains \citep{sohl2015deep, ddpm, nichol2021improved, diff_beat_gans, zhu2024thinktwiceactimproving}, and their application to astrophysical data is beginning to show considerable promise \citep{li_galaxy_evolution, Xue2023DiffusionMF, smith_galx_ddpm, lizarraga2024, jiang2024unlocking}. A critical unresolved challenge, however, lies in effectively conditioning these models on continuous physical parameters intrinsic to astrophysical systems, such as redshift.
While there have been appoaches using continuous conditioning for DDPM's such as \cite{ding2024ccdm}, or \cite{thermo}, current prevalent approaches for galaxy modeling resort to parameter discretization inevitably sacrifice information and fail to capture the continuous nature of galaxy evolution \cite{li_galaxy_evolution}.

In this study, we develop and evaluate a DDPM framework conditioned directly on continuous redshift values. We aim to demonstrate two principal advances: First, the capacity of such models to learn the conditional distribution $p(X|z)$ of galaxy images $X$ given any redshift $z$, thereby implicitly encoding the redshift-dependence of morphological characteristics. Second, we explore the model's ability under to generate galaxies conditioned at different cosmic time-points by evaluating physical metrics of the images produced. We find that the model correctly captures associated physical structure of galaxies at different time points, purely based on photometry and absent of information about these physical properties of galaxy formation, suggesting that such generative models can in fact 
produce realistic galaxies that are physically accurate to galaxies at that point in cosmic time.

As a brief review, DDPMs operate via a forward noising process and a learned reverse denoising process \cite{ho2020denoising}. The forward process gradually adds Gaussian noise $\epsilon \sim \mathcal{N}(0, I)$ to the data $X_0$ over $T$ timesteps:
\begin{equation} \label{eq:forward_ddpm_step}
 q(X_t | X_{t-1}) = \mathcal{N}(X_t; \sqrt{1-\beta_t} X_{t-1}, \beta_t I) \implies X_t = \sqrt{\bar{\alpha}_t} X_0 + \sqrt{1-\bar{\alpha}_t}\epsilon,
\end{equation}
where $\beta_t$ are predefined noise variances, $\alpha_t = 1-\beta_t$, and $\bar{\alpha}_t = \prod_{s=1}^t \alpha_s$. The model learns to reverse this process by predicting the added noise $\epsilon_\theta(X_t, t, z)$ given the noisy image $X_t$, timestep $t$, and a condition $z$ (the redshift). Generation involves iteratively sampling from $p_\theta(X_{t-1}|X_t, z)$:
\begin{equation} \label{eq:reverse_ddpm_step}
  x_{t-1} = \frac{1}{\sqrt{\alpha_t}} \left(x_t - \frac{\beta_t}{\sqrt{1-\bar{\alpha}_t}} \epsilon_\theta(X_t, t, z)\right) + \sigma_t \mathbf{z}', \quad \mathbf{z}' \sim \mathcal{N}(0,I).
\end{equation}

We focus on integrating continuous redshift $z$ as a conditioning parameter within the DDPM architecture, and leveraging the learned denoising function $\epsilon_\theta(x_t, t, z)$ to synthesize smooth morphological transitions. While Variational Autoencoders (VAEs) \citep{kingma2014autoencoding, esser2021taming} and Generative Adversarial Networks (GANs) \citep{goodfellow2014generative} have been explored for conditional generation, DDPMs offer a distinct framework with advantages in image fidelity and training stability. A core principle of our approach is to ensure that small variations in $z$ induce correspondingly subtle and physically coherent changes in the generated galaxy image $X_0$ conditioned on $z$ (denoted $X_0^z$, thereby promoting generation stability under conditioning perturbations \citep{arjovsky2017wasserstein}. 
The physical similarity of generated galaxies is rigorously assessed through quantitative morphological analysis (Section \ref{sec::evaluating_ddpm}), compared to established astrophysical observations \citep{conselice2014evolution}. Furthermore, we empirically validate the adherence of our trajectory synthesis methodology to theoretical smoothness assumptions (Section \ref{sec::evaluating_ddpm}).
\textcolor{airforceblue}{Note that redshift values near $0$ represent relatively small cosmological distances and small changes in lookback time. Where redshifts closer to $4$ correspond to much larger comoving distances. In order to keep the scale linear we require the following transform: $z' = \log(1+z)$. In this paper when we state the redshift $z$, we implicitly assume the log transform has been applied unless stated otherwise.}
\vfill

\subsection{Related Work}
\label{sec::related_work}

Generative modeling of galaxy populations has primarily employed Generative Adversarial Networks (GANs) conditioned on redshift to simulate visual characteristics \cite{lanusse2021}, \cite{margalef2020}. However, GANs are susceptible to training instabilities and mode collapse, and evaluations have often lacked rigorous morphological quantification against physical properties. While Denoising Diffusion Probabilistic Models (DDPMs) offer enhanced image quality and stability \citep{sohl2015deep, ddpm}, their application in astrophysics has largely relied on discretizing continuous parameters like redshift \citep{li_galaxy_evolution, smith_galx_ddpm}. This discretization fundamentally limits the ability to model the continuous nature of galaxy evolution and smooth morphological transitions.

The challenge of continuous conditioning is recognized across machine learning disciplines \cite{ding2024ccdm}, but robust solutions for complex astrophysical parameters like redshift, with direct physical interpretation of generated features, remain underdeveloped. Efforts to incorporate physical priors or leverage simulations \citep{galaxy_dreams, foundation_models, li_galaxy_evolution} are complementary but distinct from our data-driven approach, where evolutionary trends are learned directly from observational snapshots conditioned on continuous redshift. This limits the scope of the data the model views to limited cosmic depth, i.e. redshift values between $0$ and $1$.

\subsection{Contribution}
Our study provides the following contributions to address these limitations and advance the field:
\begin{enumerate}
    \item We establish an adaption to continuously-conditioned DDPMs, incorporating redshift perturbation during training to ensure smooth generalization across the redshift continuum. This approach overcomes the information loss associated with discretization.
    \item We demonstrate empirically that our model implicitly learns key astrophysically relevant morphological parameters (e.g., size, ellipticity, Sérsic index) and their redshift evolution, without explicit supervision from morphological labels, by solely conditioning on images and their continuous redshift values.
    \item To our knowledge, this work presents one of the first systematic constructions and evaluations of continuous galaxy evolutionary sequences from observational data using DDPMs, offering a powerful new data-driven paradigm for investigating galaxy evolution.
\end{enumerate}

\section{Dataset}
\label{sec::Data}

To date, it is prohibitively compute and time-expensive to gather redshift data from spectroscopy and pair the data to the associated photometric images. Consequently, this has severely limited the number of available datasets with matched image-redshift pairs. Among all publicly available datasets, only a few offer sufficient depth and size for training machine learning models for redshift estimation. These include the Sloan Digital Sky Survey (SDSS)~\cite{ahn2012ninth}, the Dark Energy Survey (DES)~\cite{abbott2021dark}, the Kilo-Degree Survey (KiDS)~\cite{kuijken2019kids}, and the Hyper Suprime-Cam (HSC) Survey~\cite{hyper_data}.

\begin{figure}
    \centering
    \includegraphics[width=0.5\linewidth]{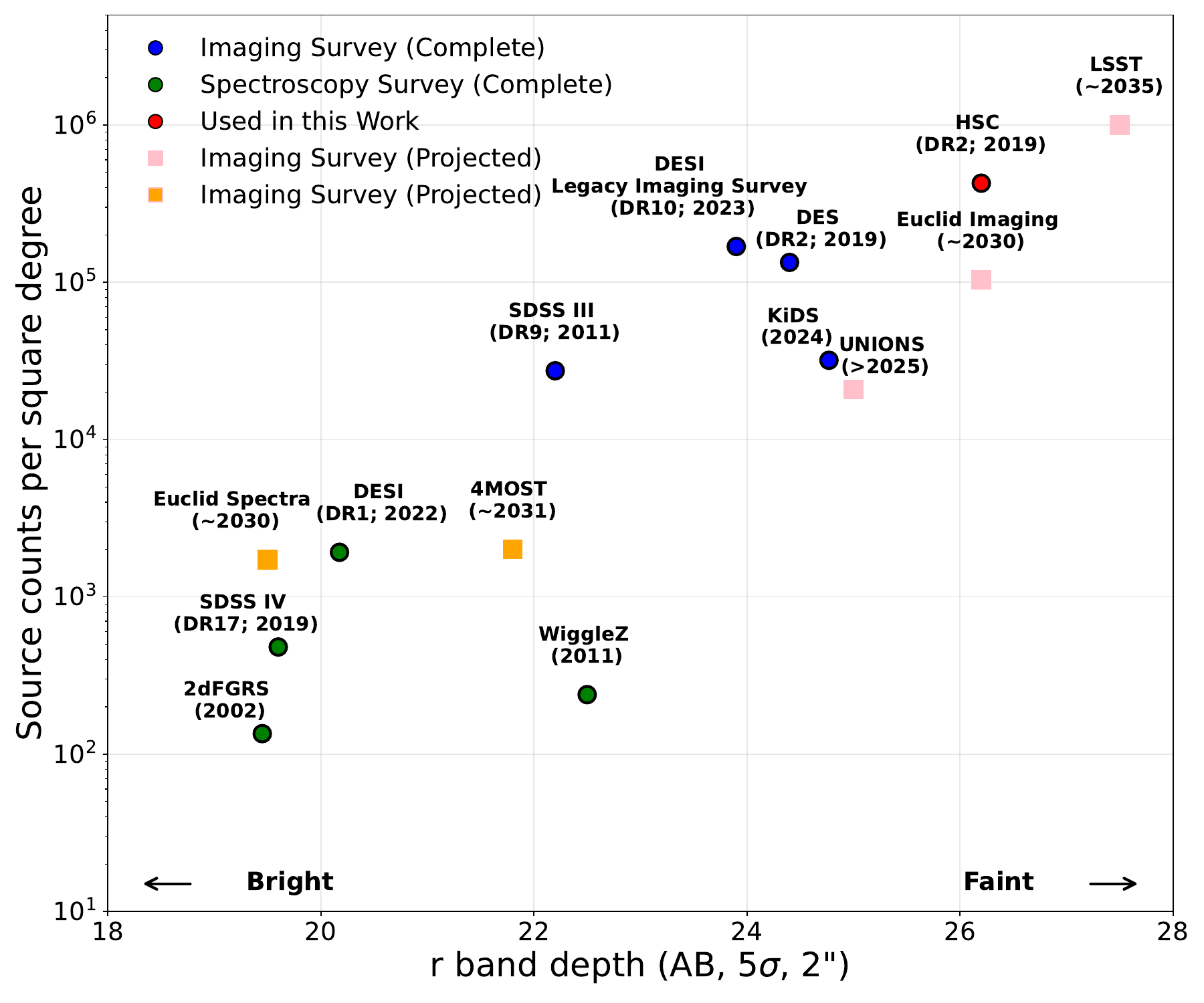} 
    \label{fig:number_density_depth}
\caption{
Source density versus $r$-band depth for major imaging and spectroscopic galaxy surveys. The $r$-band depth corresponds to the $5\sigma$ detection limit in a $2''$ aperture. Each point represents a survey, with completed imaging surveys (e.g., SDSS, DES, KiDS, HSC) shown in solid markers, and projected surveys (e.g., LSST, Euclid) shown in pink outlined squares. The y-axis shows the number of detected sources per square degree on a logarithmic scale. Among publicly available datasets, the Hyper Suprime-Cam (HSC) survey stands out as the deepest complete imaging survey, with a source density of $\sim 10^5$ galaxies per deg$^2$ and an $r$-band depth of $\sim 26.5$ AB magnitudes. These properties make HSC uniquely suited for training deep learning models that require high-resolution, redshift-labeled galaxy images.
}
\end{figure}

We employ the \textbf{Hyper Suprime-Cam Galaxy Dataset} curated by Do et al.~\citet{do_2024_11117528}, publicly accessible at Zenodo (GalaxiesML: \url{https://zenodo.org/records/11117528}, CC-BY 4.0). HSC is the deepest of the completed surveys, with high-quality redshift–image pairs. As shown in Figure~\ref{fig:number_density_depth}, HSC achieves an $r$-band depth of $\sim26.5$ AB magnitudes and a source density of $\sim 10^5$ galaxies per square degree—an order of magnitude higher than SDSS and DES. Other upcoming surveys like LSST and Euclid are projected to reach similar or greater depths (shown as pink squares), but their data is not yet publicly released.
The dataset comprises 286,401 galaxies, each with spectroscopic redshift measurements spanning $z = 0$ to $z = 4$. Redshift indicates a galaxy’s distance and the age of the universe when its light was emitted; for example, $z = 1$ corresponds to light emitted approximately 7.8 billion years ago.
Each galaxy is represented by a $64 \times 64$ pixel image in five optical wavelength bands: $g, r, i, z, y$. These filters correspond to progressively redder parts of the visible and near-infrared spectrum. For context, $g$ captures green light at $\sim477$~nm, while $y$ captures near-infrared light around $\sim1~\mu$m.
Due to selection effects in spectroscopic follow-up, the dataset is biased toward lower redshifts, with about 92.8\% of galaxies having $z < 1.5$ (corresponding to light emitted within the past $\sim$9.5 billion years). We follow the training (204,513 images) and testing (40,914 images) split proposed by Li et al.~\cite{li2024unredshift}.


\section{Methodology}
\label{sec::methods}
\subsection{Continuous Redshift-Conditioned DDPM Architecture}
Utilizing DDPMs \cite{ddpm}, we introduce a novel approach to learn the conditional distribution \( p(X^z \mid z) \) by integrating redshift values into the U-Net architecture's time steps \cite{li_galaxy_evolution, smith_galx_ddpm}. To prevent model overfitting and ensure learning is concentrated within a Gaussian neighborhood around specific redshifts \( z \), Gaussian noise \( \mathcal{N}(0, \sigma^2) \) is added to the redshifts during training, enhancing the model's ability to interpolate between nearby redshifts. Our Conditional Denoising U-Net starts with a noisy initial galaxy image  \( X_T^z \) and, through iterative denoising informed by both time step and the adjusted redshifts, aims to produce a clean galaxy image  \( X_0^z \). To additionally stabilize the training, we implement an Exponential Moving Average (EMA) \cite{EMA_Karras2024edm2} and adhere to a standard variance schedule \cite{ddpm, ddim} to balance noise addition and preserve data structure.

\begin{figure}
    \centering
    \includegraphics[width=0.5\linewidth]{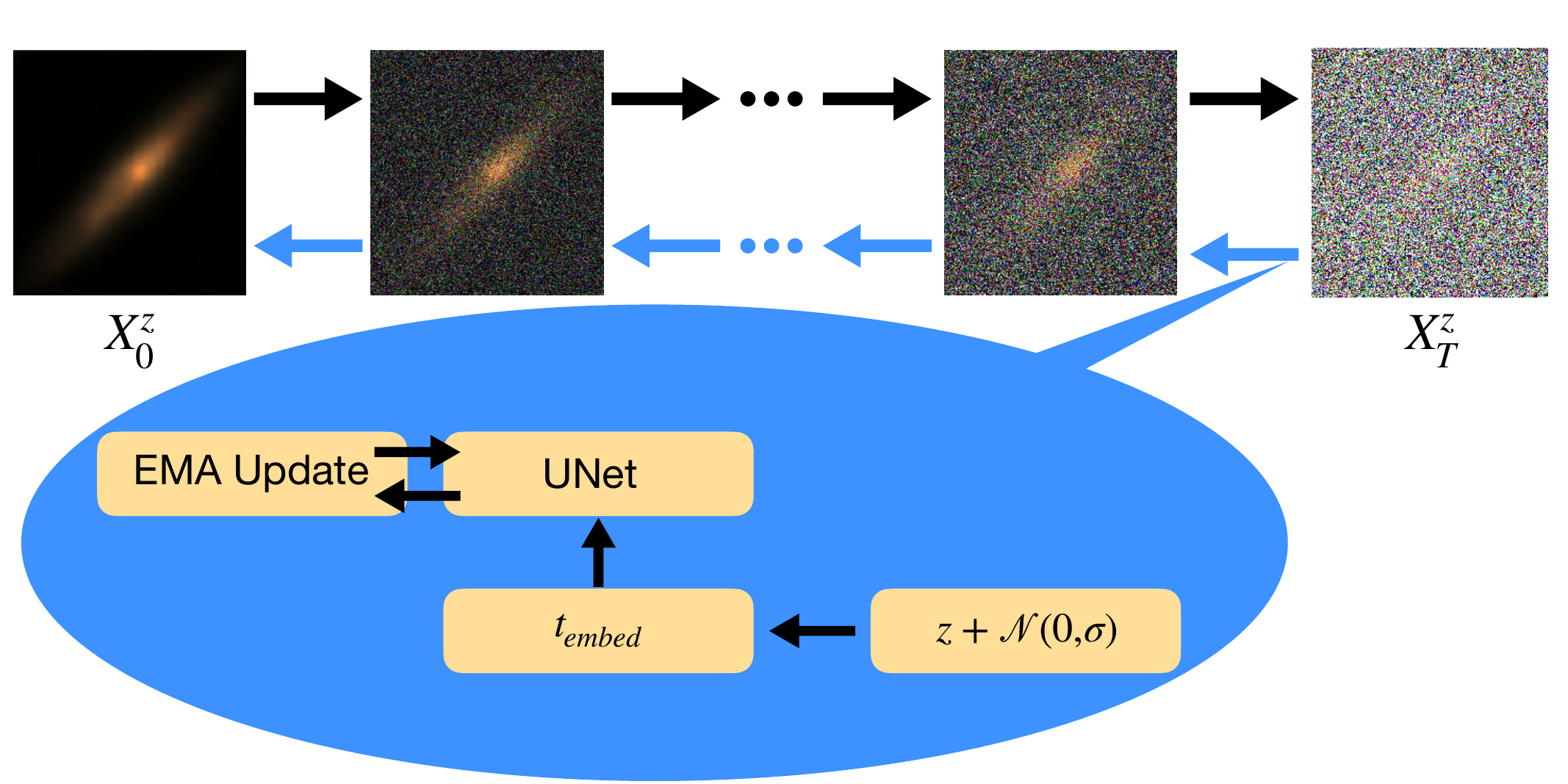} 
    \label{fig:ddpm_architecture}
\caption{Model Architecture: Our model follows conventional DDPM implementations, but the noise adjust conditioning of $z$ allows the model to better interpolate between it's conditioning for nearby neighborhoods of $z$.}
\end{figure}

The model's diffusion process starts with \(64 \times 64\) pixel galaxies images with 5 channels, which are passed to a noising schedule across $1,000$ time steps, linearly interpolating noise levels from a Beta Start of \(1 \times 10^{-4}\) to a Beta End of 0.02. Training utilizes Huber Loss for its robustness to outliers, gradient clipping with a maximum norm of 1.0, and an AdamW optimizer \cite{loshchilov2017decoupled} set to a learning rate of \(2 \times 10^{-5}\). Redshifts are perturbed with Gaussian noise (std dev 0.01) to prevent overfitting and improve generalization. Our U-Net model, equipped with self-attention layers, varies channels by resolution stage and includes 4 attention heads with layer normalization and GELU activation \cite{hendrycks2016gaussian}, applied before and after attention. Temporal and conditional redshift information is encoded using sinusoidal positional encoding of the time step \(t\), transformed into a 256-dimensional vector. This vector is further modified by adding Gaussian noise to the redshift value \(z + \mathcal{N}(0,0.01)\), prior to being fed into the U-Net Fig. \ref{fig:ddpm_architecture}. The model was trained on a single NVIDIA A6000 GPU.

\section{Evaluation}
\label{sec::evaluating_ddpm}

\begin{figure} 
 \centering
 \includegraphics[width=0.3\linewidth]{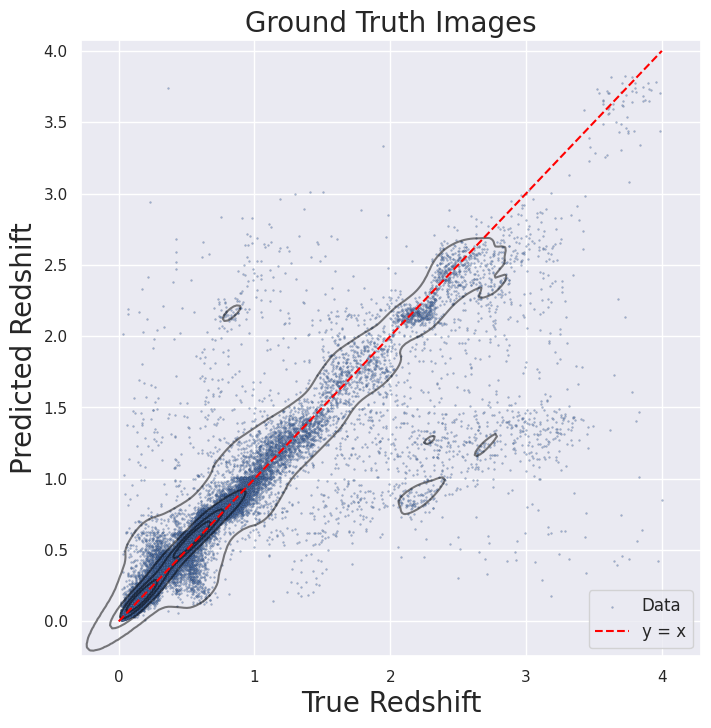} 
 \includegraphics[width=0.3\linewidth]{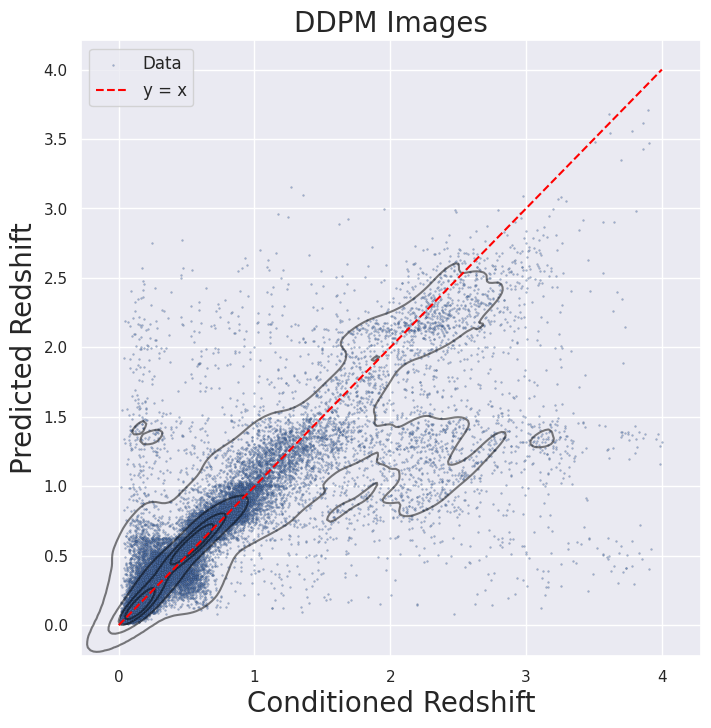} 
 \includegraphics[width=0.3\linewidth]{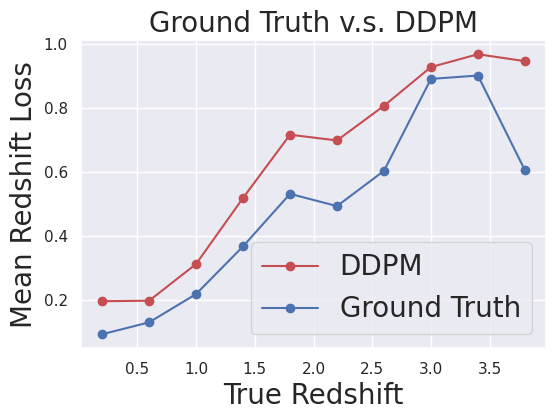} 
 \caption{Redshift prediction quality for synthesized galaxies. (Left) CNN-predicted redshift $\hat{z}$ versus true spectroscopic redshift $z$ for real test-set galaxies, demonstrating the baseline accuracy of the independent predictor. (Middle) CNN-predicted $\hat{z}$ versus the input conditioning redshift $z$ for DDPM-generated galaxies, showing strong correlation. (Right) Mean redshift loss, $|\hat{z}-z|/(1+z)$, as a function of true/conditioned $z$, confirming good performance in well-sampled regions (e.g., $z < 1.5-2.0$) and highlighting increased scatter at higher redshifts due to training data sparsity.}
 \label{fig::ground_vs_real}
\end{figure}

Our evaluation aims to quantitatively and qualitatively assess the capacity of our redshift-conditioned DDPM and compare it against recently proposed architectures. The primary objectives were to determine its efficacy in learning meaningful physical and morphological characteristics of galaxies from image data alone, and its ability to generate physically plausible evolutionary sequences across cosmic time. In particular, we focus on the measured physical attributes of the HyperCam-Suprime survey to gauge the physical consistency of our generated images, which involve five color filters $(g, r, i, z, y)$. \textcolor{airforceblue}{While perceptual quality metrics like Fréchet Inception Distance (FID) \cite{heusel2017gans} indicate general similarity to true images, they fail to assess critical morphological properties of galaxies and their evolution over time. It's also important to note that the original VGG-Net used in \cite{heusel2017gans}, was trained on $3$-channeled, $256$ pixeled RGB images. To compute FID, \cite{li_galaxy_evolution, lanusse2021}, propose sub-sampling the color filters to $(g,r,i)$. We follow suit with this in Table \ref{tab::benchmarks}, however we suspect that this is quite a dubious comparison, as we don't compare all the color channels. Moreso, the range of these images is continuous and much larger than that of natural RGB images.} 
To aid our evaluation, we generate synthetic images conditioned on redshifts from the test dataset and compare their physical properties that astronomers typically use to characterize galaxies \cite[e.g.][]{conselice2014}, such as the shape (ellipticity, semi-major axis), size (isophotal area), and brightness distribution (Sérsic index). Furthermore, using the CNNRedshift predictor established by Li et al. \cite{li_galaxy_evolution}, we assess the redshift accuracy against the ground truth, utilizing the redshift loss from \cite{nishizawa2020photometricredshiftshypersuprimecam}. This redshift predictor was trained on real galaxy images using spectroscopic ground truth and produces good predictions on real data (Fig. \ref{fig::ground_vs_real}). These comparisons help verify the physical plausibility of the diffusion model's output.
We compute standard morphological metrics for both the test data and the DDPM-generated images conditioned on the test data's redshifts:

\begin{itemize}
    \item \textbf{Ellipticity}: A measure of how elongated a galaxy appears, calculated as $\epsilon = 1 - \frac{b}{a}$, where $a$ and $b$ are the semi-major and semi-minor axes, respectively.
    
    \item \textbf{Semi-major Axis}: The length of the longest axis of the galaxy's elliptical shape, indicating its size.
    
    \item \textbf{Sérsic Index}: Describes the intensity profile \( I(r) \) of a galaxy as a function of radius \( r \), given by:
    \[
    I(r) = I_0 \exp\left( -b_n \left( \left( \frac{r}{r_e} \right)^{1/n} - 1 \right) \right),
    \]
    where \( I_0 \) is the intensity at \( r = 0 \), \( r_e \) is the effective radius, \( n \) is the Sérsic index, and \( b_n \) is a constant dependent on \( n \). Higher Sérsic indices indicate more concentrated light profiles.
    
    \item \textbf{Isophotal Area}: The area over which the galaxy emits light above a certain intensity threshold, reflecting the galaxy's apparent size.
\end{itemize}

\begin{figure}
    \centering
    \includegraphics[width=1.0\linewidth]{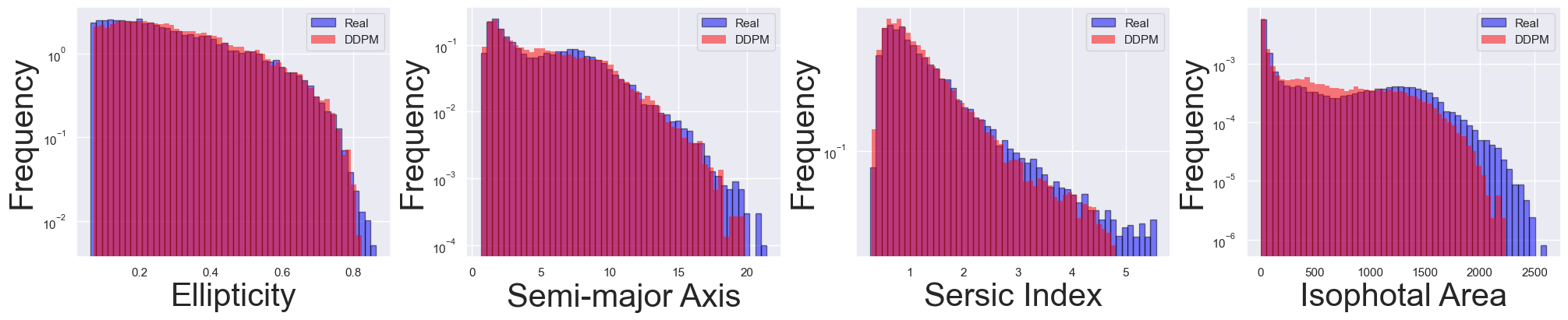} 
    \caption{
    From left to right, the figure displays histograms comparing the frequency distribution of DDPM-generated and real galaxies in terms of 1) ellipticity, 2) semi-major axis,  3) Sérsic index, and 4 isophotal area in the log-scale of redshift $z$.
    }
    \label{fig::morph_hist}
\end{figure}

Our findings confirm that the DDPM successfully learns the physical characteristics of galaxies even though these attributes were not explicitly provided during training. When comparing the distributions of each metric between the DDPM-generated images and the real data, we observe that the overall shapes of the distributions are very similar, as shown in Figure \ref{fig::morph_hist}. This indicates that for any conditioned redshift, the model produces physically plausible galaxy images that reflect the morphological diversity present in the real dataset.

\begin{figure}
    \centering
    \includegraphics[width=1.0\linewidth]{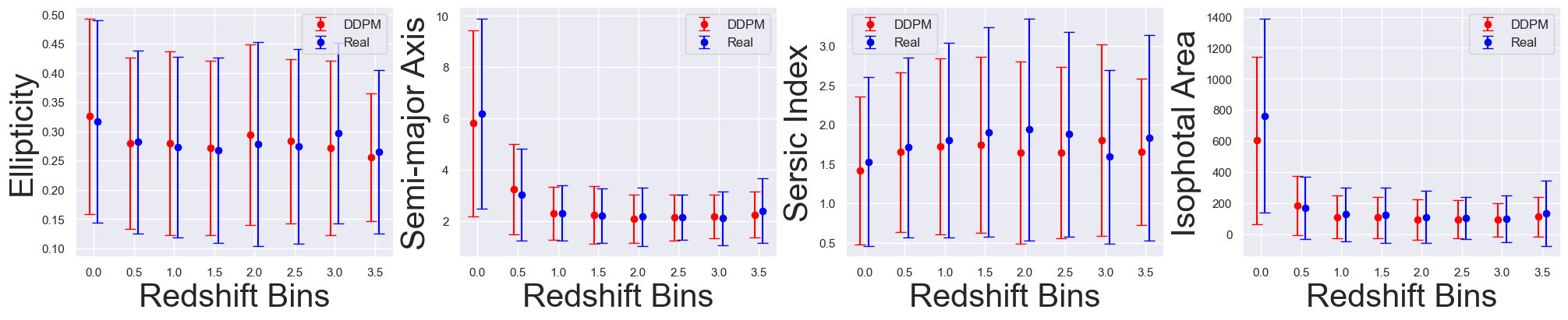} 
    \caption{Mean morphological metrics as a function of redshift. Comparison between real test galaxies (orange) and DDPM-generated galaxies (blue) across different redshift bins, with error bars representing $95\%$ confidence intervals on the mean. The model accurately reproduces the observed evolutionary trends in average ellipticity, semi-major axis, Sérsic index, and isophotal area with redshift.}
    \label{fig::morph_bars}
\end{figure}

In Figure \ref{fig::morph_bars}, we show that for each redshift bin, the mean values of these metrics for DDPM-generated galaxies closely match those of the real test distribution. The error bars represent the $95\%$ confidence intervals, indicating that the generated images exhibit similar variability to the real data. This suggests that the DDPM model is able to associate redshifts with morphological characteristics of galaxies observed at that redshift, effectively capturing the trends in galaxy evolution.

Although the redshift predictions for generated images show increased variance at higher redshifts (see Figure \ref{fig::ground_vs_real}), the morphological characteristics remain physically plausible (Figures \ref{fig::morph_hist} and \ref{fig::morph_bars}). This indicates that while the model captures the general trends in galaxy morphology with redshift, it may produce a broader range of images at higher redshift values, potentially blending characteristics from neighboring redshifts due to data scarcity in those regions. This effect is evident in Figure \ref{fig::real_vs_ddpm_visual}, where the generated images display increased diversity and variability at higher redshifts.

\subsection{Benchmarks}
\label{sec:benchmarks}

There are few generative models trained on deep red-shift survey's (e.g. HyperCam-Suprime). In order to bechmark and compare this model to similar models, we compare against the GAN in \cite{margalef2020} and the DDPM implemented by \cite{smith_galx_ddpm}. Since neither of these models were directly using redshift-conditioning, we re-implement these architectures according to their specifications and train continuous red-shift conditioned models using the Guassian perturbation mentioned in Sec. \ref{sec::methods} binned, discrete red-shift conditioned versions of the model the performance in terms of FID, Ellipticity, Semi-Major Axis, Sérsic Index, and Isophotal Area can be found in Table \ref{tab::benchmarks}. Additionally, we ablate on $\sigma$ for the continuous conditioned models and find that our addpation of the DDPM benefits the most from lower values of $\sigma$ but when $\sigma = 0$ (i.e. discrete conditioning DDPM), the physical measurements take a sharp drop in performance even though the FID is lower. Note we sample $10,000$ redshift values $z$, from the test set and compute their average Ellipticity, Semi-Major Axis, ect. Then, we take each model synthesize images across all these models and compute the FID (on the false image using the sub-sampling of channels mentioned in earlier), and we compute the average morphological metrics. A lower FID score indicate perceptually similar images, and the other physical metrics are taken as ratios between the average computed in the synthetic data divided by the average computes in the real test data (a score closer to 1 is better). It's worth noting that our model achieve the second best FID score our of the model that were compared, however, our model has notably better performance in terms of the physical benchmarks that astronomers typically use. This further supports our hypothesis that conditioning DDPM on redshift is sufficient for the model to implicitly understand some physical properties of galaxies at certain redshift levels. Moreso, being able to appropriately embed the redshift value has a dramatic effect on the model ability to capture physical properties.

\begin{table} 
\label{tab::benchmarks}
\centering
\resizebox{\textwidth}{!}{%
\begin{tabular}{lccccc}
\toprule
\textbf{Model} & \textbf{FID} $\downarrow$ & \textbf{Ellipticity} $\%$ & \textbf{Semi-Major Axis} $\%$ & \textbf{Sérsic Index} $\%$ & \textbf{Isophotal Area} $\%$\\
\midrule
$z_{\text{discrete}}$-cond. GAN \cite{margalef2020}$^\star$                    & 28.7          & 0.07          & 0.04          & 0.61          & 0.48 \\
$z_{\text{discrete}}$-cond. DDPM \cite{smith_galx_ddpm}$^\star$                & \textbf{11.6} & 0.45          & 0.22          & 0.38          & 0.29 \\
$z_{\text{continuous}}$-cond. GAN $\sigma = 1.0$ \cite{margalef2020}$^\star$   & 45.0          & 0.12          & 0.11          & 0.68          & 0.40 \\
$z_{\text{continuous}}$-cond. GAN $\sigma = 0.5$ \cite{margalef2020}$^\star$   & 30.3          & 0.11          & 0.17          & 0.48          & 0.39 \\
$z_{\text{continuous}}$-cond. GAN $\sigma = 0.1$ \cite{margalef2020}$^\star$   & 25.3          & 0.15          & 0.20          & 0.58          & 0.43 \\
$z_{\text{continuous}}$-cond. DDPM (ours) $\sigma = 1.0$                     & 24.0          & 0.28          & 0.50          & 0.40          & 0.20 \\
$z_{\text{continuous}}$-cond. DDPM (ours) $\sigma = 0.5$                     & 15.2          & 0.82          & 0.70          & 0.81          & 0.76 \\
$z_{\text{continuous}}$-cond. DDPM (ours) $\sigma = 0.1$                     & 14.2          & \textbf{0.98} & \textbf{0.96} & \textbf{0.93} & \textbf{0.90} \\
\bottomrule
\end{tabular}%
}
\caption{Quantitative metrics for generative models of galaxy images conditioned on redshift. Lower FID is better; other columns are measured in terms of the proportion of the physical measurements compared to the true galaxy data-test-set measurements (closer to 1 indicates nearly perfect accuracy). Notice that the FID alone is not a good indication if the model is suitable for scientific studies of galaxies. Row 2 has the lowest FID, but the physical measurements of the galaxies produced don't accurately represent physical galaxy characteristics across the redshift values in the test set, despite appearing more realistic.}
\label{tab:galaxy_gen_metrics}
\noindent\textsuperscript{*}\parbox[t]{\textwidth}{\footnotesize Indicates that the original model was not trained on the HyperCam-Suprime survey. We reimplemented and retrained these models with redshift conditioning to keep the comparison as fair as possible.}

\end{table}

\subsection{Visual Comparisons and Analysis}

\begin{figure}
    \centering
    \includegraphics[width = 1.0\linewidth]{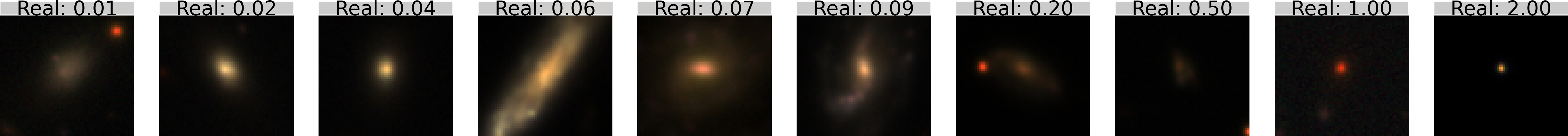} 
    \includegraphics[width = 1.0\linewidth]{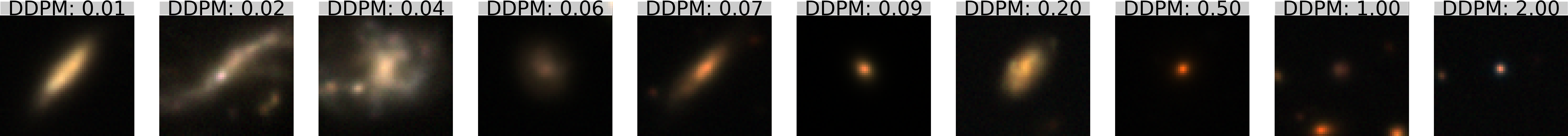} 
    \caption{Visual comparison of real and synthesized galaxies across redshifts. (Top row) Examples of real galaxies from the Hyper Suprime-Cam test set at various spectroscopic redshifts. (Bottom row) DDPM-generated galaxies conditioned on the same respective redshifts as the real examples above them. The model successfully captures redshift-dependent visual characteristics, including changes in apparent size, color, and structural definition.}
    \label{fig::real_vs_ddpm_visual}
\end{figure}

Visual comparisons between real and generated galaxy images further illustrate the model's ability to capture the morphological characteristics associated with different redshifts. Figure \ref{fig::real_vs_ddpm_visual} presents side-by-side images of real galaxies and DDPM-generated galaxies conditioned on the same redshifts. The generated images exhibit realistic galaxy features, capturing variations in structure, size, and brightness distribution that are consistent with changes in redshift. \textcolor{airforceblue}{Caution: The RGB images of the galaxies presented in this paper uses (g,r,i) to make the false color image as recommnended in \cite{li_galaxy_evolution}. Consequently, the $i$ and $r$ channels can pick up heat and dust and project it in the image, making the image appear noise. As a consequence, some of the DDPM generated images will also have this as an artifact in when projecting the output to a False color.}

As redshift increases, galaxies tend to appear smaller and less defined due to cosmological effects such as cosmic expansion and the redshifting of light. Our model reflects these trends, indicating that it has learned meaningful relationships between redshift and galaxy morphology, despite only being conditioned on redshift and not explicitly provided with morphological labels during training.

These results suggest that our model not only generates visually plausible galaxy images but also accurately reflects the underlying physical properties associated with different redshifts. The ability of the DDPM to implicitly capture galaxy morphology demonstrates the effectiveness of continuous conditioning in modeling complex, continuous variations in data, such as galaxy evolution over cosmic time.

\begin{figure}
    \centering
     \includegraphics[width =1.0\linewidth]{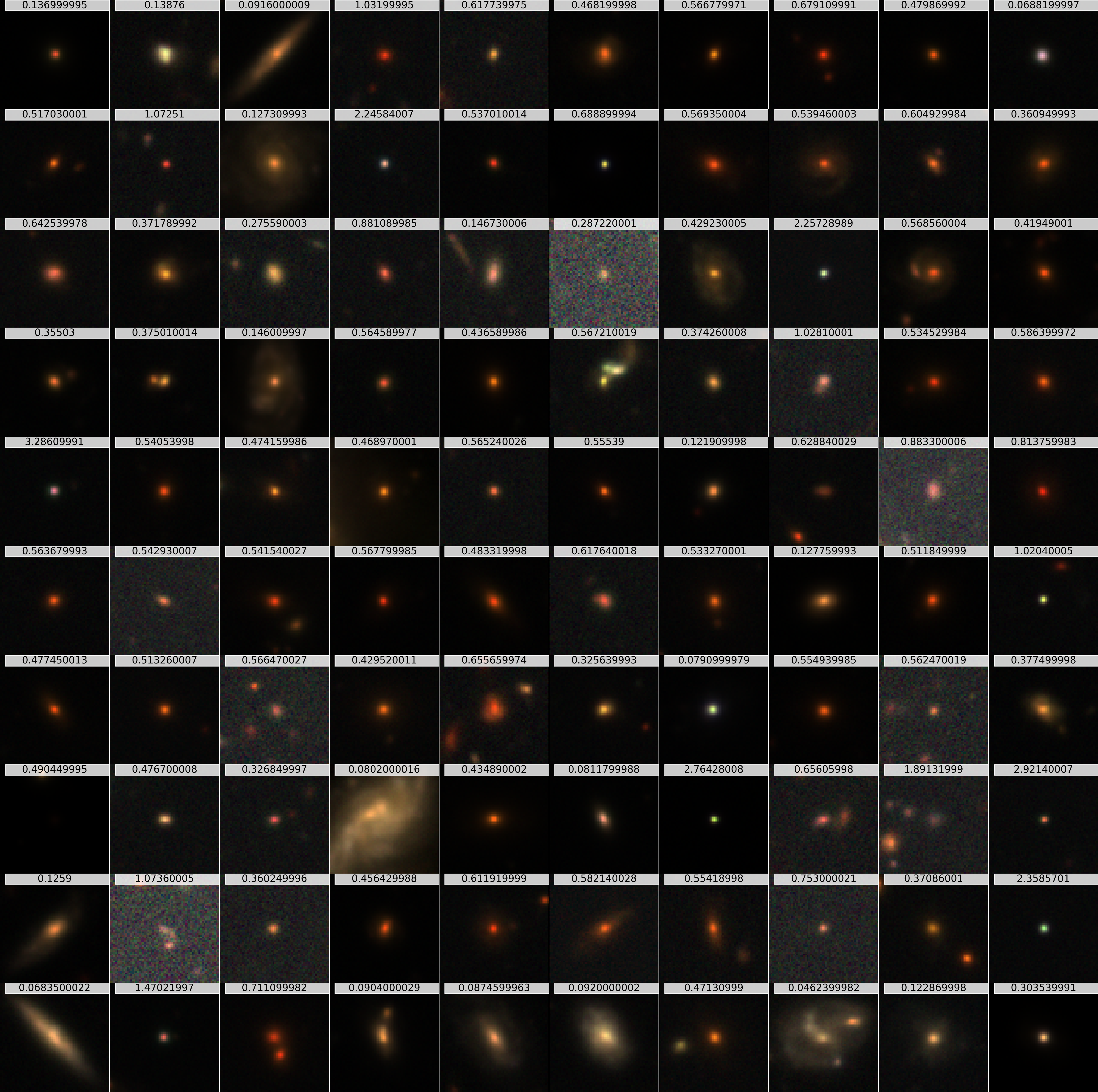}
    \caption{DDPM (False Color) Galaxies Generated at non-cherry-picked redshift values.}
\end{figure}

\section{Discussion and Limitations}
\label{sec::discussion_limitations}
Despite the promising results of our model, several limitations need to be acknowledged. One of the key challenges is that galaxies do not evolve in isolation. Our model currently treats each galaxy independently, failing to account for the complex interactions between galaxies and their environments, such as mergers or gravitational interactions. These interactions play a significant role in galaxy evolution, and ignoring them may limit the physical when applying it to clusters of galaxies.

\begin{figure}
    \centering
    \includegraphics[width = 0.9\linewidth]{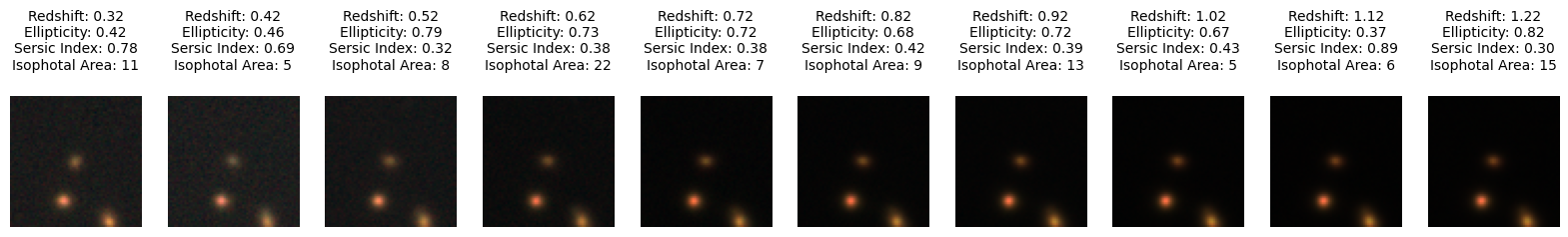}
    \includegraphics[width = 0.9\linewidth]{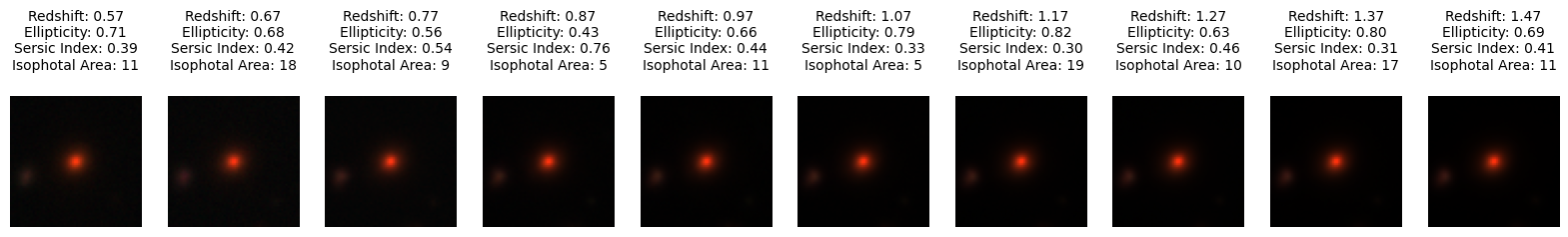}
    \includegraphics[width = 0.9\linewidth]{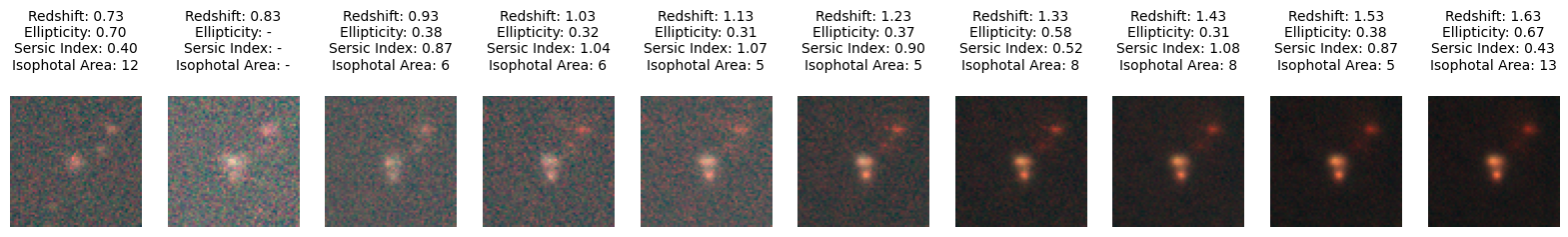}   
    \caption{Real images and their corresponding denoised outputs for pertubed $z + \Delta z$. Viewing physical metrics as the we construct a trajectory based on the real image (Left). The model is only conditioned on redshift, while metrics such as Sérsic Index, Ellipticity, Isophotal area may be assocated with certain redshift bins as in Fig. \ref{fig::morph_hist}, the variance of 
    these metrics can be higher for these trajectories which are out of distribution. Note that depending on noise-to-signal, some metrics computations are not computed and are left blank.}
    \label{fig::img_traj_metrics}
\end{figure}

Additionally, while our model successfully generates realistic galaxy images conditioned on redshift, the denoising process might inadvertently remove noise that encodes important physical information in later stages of galaxy evolution. This smoothing effect could reduce the overall astrophysical validity of the generated data, particularly when simulating high-redshift galaxies. Additionally metrics such as Sérsic index, ellipticity, and isophotal area can potentially have higher variance when perturbing the redshift too much. To understand this, we took a random sample real images from the test dataset. On the left of Fig. \ref{fig::img_traj_metrics} is the initial image. Note that same galaxies appear to be noisy, but this is merely dust and residual heat being captured in the image.
Since the conditioning of the DDPM was on $z$ with a Gaussian perturbation, we feed the image in the model and denoise it at $z + \Delta z$, where $\Delta z = 0.1$. Notice that same of the galaxies stay mostly the same (with a slightly more red hue towards the end), but row $3$ has dramatic changes.

Moreover, the model's performance is notably less reliable at higher redshifts, where the training data is sparse. This limitation indicates that the model struggles to capture the full diversity of galaxy morphologies at these redshifts, leading to increased variability in the generated images and less accurate redshift conditioning. To make matters more challenging, other scientific domains have found that model performance may not capture scientifically relevant structure even in simpler models such as VAE's~\cite{pmlr-v194-lizarraga22a, Nunez_2021_CVPR, streamline}, 
VQ-VAE's~\cite{isbi}, and MLP's~\cite{monitoring, honig2025betterpromptcompressionmultilayer} further amplifying the challenge of correctly interpreting the results of more sophisticaged architectures like a DDPM.

\section{Conclusion}
In this work, we introduced a novel approach to generating galaxy images using Denoising Diffusion Probabilistic Model (DDPM), conditioned on continuous redshift values. Our results show that the DDPM effectively captures essential physical attributes of galaxies, such as semi-major axis, isophotal area, ellipticity, and Sérsic index, with high fidelity to the true data distribution. This suggests that redshift, a measure of both age and distance, serves as a strong predictor of galaxy structure, even without direct morphological inputs. 

Future work should focus on extending this approach toward models that can learn the physical evolution of galaxies more directly. Reproducing the morphological characteristics \cite[e.g.,][]{conselice2014} is the first step to embed the physics of galaxy evolution into an neural network. To demonstrate more direct connection to physics, one should also apply more stringent tests. For example, it's uncertain if models produce galaxies that have the same star formation rate density evolution \cite[e.g.,][]{madau2014} or physical changes through galaxy mergers \cite{lotz2008}.

Moreover, considering DDPM's ability to interpolate between modes of the learned probability distribution, we propose raises the question if DDPM's can be utilized for dynamic visualizations of galaxy evolution as a function of redshift. Such a framework could serve as a powerful tool for studying galaxy formation and evolution across cosmic timescales.

\subsection{Future Work} There are few auto-regressive (LLM-based) models on galaxy data. One main concern as mentioned earlier the amount of data in survey's such as the HyperCam-Suprime dataset are very low for auto-regressive models. However a new line of LLM models with latent cross-attention mechanisms have been shown to work well with low-data bottle necks and offer interpretability due to the latent-prompt learning the heirarchical structure of the data through a meta-learning process~\cite{latent-RL, lizarraga2025rlisntenough}. Future work can explore these directions further.

\begin{ack}

This work was partially supported by CSREx - SOU, NSF DMS-2015577, NSF DMS-2415226, NSF-DGE-2034835 and a gift fund from Amazon.
\end{ack}

\bibliographystyle{plainnat}
\bibliography{ref}

\begin{thebibliography}{47}
\providecommand{\natexlab}[1]{#1}
\providecommand{\url}[1]{\texttt{#1}}
\expandafter\ifx\csname urlstyle\endcsname\relax
  \providecommand{\doi}[1]{doi: #1}\else
  \providecommand{\doi}{doi: \begingroup \urlstyle{rm}\Url}\fi

\bibitem[Abbott et~al.(2022)]{abbott2021dark}
T.~M.~C. Abbott et~al.
\newblock Dark energy survey year 3 results: Cosmological constraints from galaxy clustering and weak lensing.
\newblock \emph{Physical Review D}, 105\penalty0 (2):\penalty0 023520, 2022.

\bibitem[Ahn et~al.(2012)]{ahn2012ninth}
C.~P. Ahn et~al.
\newblock The ninth data release of the sloan digital sky survey: First spectroscopic data from the sdss-iii baryon oscillation spectroscopic survey.
\newblock \emph{The Astrophysical Journal Supplement Series}, 203\penalty0 (2):\penalty0 21, 2012.

\bibitem[Aihara et~al.(2019)]{aihara2019second}
H.~Aihara et~al.
\newblock Second data release of the hyper suprime-cam subaru strategic program.
\newblock \emph{Publications of the Astronomical Society of Japan}, 71\penalty0 (5):\penalty0 114, 2019.

\bibitem[Arjovsky et~al.(2017)Arjovsky, Chintala, and Bottou]{arjovsky2017wasserstein}
Martin Arjovsky, Soumith Chintala, and L{\'e}on Bottou.
\newblock Wasserstein gan.
\newblock In \emph{International Conference on Machine Learning (ICML)}, pages 214--223, 2017.

\bibitem[Conselice(2014{\natexlab{a}})]{conselice2014}
Christopher~J. Conselice.
\newblock The {{Evolution}} of {{Galaxy Structure Over Cosmic Time}}.
\newblock \emph{Annual Review of Astronomy and Astrophysics}, 52\penalty0 (1):\penalty0 291--337, 2014{\natexlab{a}}.
\newblock \doi{10.1146/annurev-astro-081913-040037}.

\bibitem[Conselice(2014{\natexlab{b}})]{conselice2014evolution}
Christopher~J Conselice.
\newblock The evolution of galaxy structure over cosmic time.
\newblock \emph{Annual Review of Astronomy and Astrophysics}, 52:\penalty0 291--337, 2014{\natexlab{b}}.

\bibitem[Dhariwal and Nichol(2021)]{diff_beat_gans}
Prafulla Dhariwal and Alexander Nichol.
\newblock {Diffusion Models Beat GANs on Image Synthesis}.
\newblock In M.~Ranzato, A.~Beygelzimer, Y.~Dauphin, P.S. Liang, and J.~Wortman Vaughan, editors, \emph{Advances in Neural Information Processing Systems}, volume~34, pages 8780--8794. Curran Associates, Inc., 2021.
\newblock URL \url{https://proceedings.neurips.cc/paper_files/paper/2021/file/49ad23d1ec9fa4bd8d77d02681df5cfa-Paper.pdf}.

\bibitem[Ding et~al.(2024)Ding, Wang, Zhang, and Wang]{ding2024ccdm}
Xin Ding, Yongwei Wang, Kao Zhang, and Z.~Jane Wang.
\newblock {CCDM}: Continuous conditional diffusion models for image generation, 2024.

\bibitem[Do et~al.(2024{\natexlab{a}})Do, Lizarraga, Yeh, et~al.]{do2024galaxiesml}
Quynh Do, Andrew Lizarraga, Claudia Yeh, et~al.
\newblock Galaxiesml: A dataset of galaxy images with matched redshifts.
\newblock \emph{Zenodo}, 2024{\natexlab{a}}.
\newblock URL \url{https://zenodo.org/records/11117528}.

\bibitem[Do et~al.(2024{\natexlab{b}})Do, Jones, Boscoe, Li, and Alfaro]{do_2024_11117528}
Tuan Do, Evan Jones, Bernie Boscoe, Yunqi~(Billy) Li, and Kevin Alfaro.
\newblock {GalaxiesML: an imaging and photometric dataset of galaxies for machine learning}, June 2024{\natexlab{b}}.
\newblock URL \url{https://doi.org/10.5281/zenodo.11117528}.

\bibitem[Esser et~al.(2021)Esser, Rombach, and Ommer]{esser2021taming}
Patrick Esser, Robin Rombach, and Bj{\"o}rn Ommer.
\newblock Taming transformers for high-resolution image synthesis.
\newblock \emph{Proceedings of the IEEE/CVF Conference on Computer Vision and Pattern Recognition (CVPR)}, pages 12873--12883, 2021.

\bibitem[Goodfellow et~al.(2014)Goodfellow, Pouget-Abadie, Mirza, Xu, Warde-Farley, Ozair, Courville, and Bengio]{goodfellow2014generative}
Ian~J. Goodfellow, Jean Pouget-Abadie, Mehdi Mirza, Bing Xu, David Warde-Farley, Sherjil Ozair, Aaron Courville, and Yoshua Bengio.
\newblock Generative adversarial nets.
\newblock In Z.~Ghahramani, M.~Welling, C.~Cortes, N.~Lawrence, and K.Q. Weinberger, editors, \emph{Advances in Neural Information Processing Systems}, volume~27. Curran Associates, Inc., 2014.
\newblock URL \url{https://proceedings.neurips.cc/paper_files/paper/2014/file/f033ed80deb0234979a61f95710dbe25-Paper.pdf}.

\bibitem[Hendrycks and Gimpel(2016)]{hendrycks2016gaussian}
Dan Hendrycks and Kevin Gimpel.
\newblock Gaussian error linear units (gelus).
\newblock \emph{arXiv preprint arXiv:1606.08415}, 2016.

\bibitem[Heusel et~al.(2017)Heusel, Ramsauer, Unterthiner, Nessler, and Hochreiter]{heusel2017gans}
Martin Heusel, Hubert Ramsauer, Thomas Unterthiner, Bernhard Nessler, and Sepp Hochreiter.
\newblock {GANs Trained by a Two Time-Scale Update Rule Converge to a Local Nash Equilibrium}.
\newblock In \emph{Advances in Neural Information Processing Systems (NeurIPS)}, volume~30, pages 6626--6637, 2017.

\bibitem[Hiroaki~Aihara and et~al.(2019)]{hyper_data}
Makoto~Ando Hiroaki~Aihara, Yusra~AlSayyad and et~al.
\newblock {{Second data release of the Hyper Suprime-Cam Subaru Strategic Program}}.
\newblock \emph{Publications of the Astronomical Society of Japan}, 71\penalty0 (6):\penalty0 114, 10 2019.
\newblock ISSN 0004-6264.
\newblock \doi{10.1093/pasj/psz103}.
\newblock URL \url{https://doi.org/10.1093/pasj/psz103}.

\bibitem[Ho et~al.(2020{\natexlab{a}})Ho, Jain, and Abbeel]{ddpm}
Jonathan Ho, Ajay Jain, and Pieter Abbeel.
\newblock {Denoising Diffusion Probabilistic Models}.
\newblock In H.~Larochelle, M.~Ranzato, R.~Hadsell, M.F. Balcan, and H.~Lin, editors, \emph{Advances in Neural Information Processing Systems}, volume~33, pages 6840--6851. Curran Associates, Inc., 2020{\natexlab{a}}.
\newblock URL \url{https://proceedings.neurips.cc/paper_files/paper/2020/file/4c5bcfec8584af0d967f1ab10179ca4b-Paper.pdf}.

\bibitem[Ho et~al.(2020{\natexlab{b}})Ho, Jain, and Abbeel]{ho2020denoising}
Jonathan Ho, Ajay Jain, and Pieter Abbeel.
\newblock Denoising diffusion probabilistic models.
\newblock In \emph{Advances in Neural Information Processing Systems (NeurIPS)}, pages 6840--6851, 2020{\natexlab{b}}.

\bibitem[Honig et~al.(2025)Honig, Lizarraga, Zhang, and Wu]{honig2025betterpromptcompressionmultilayer}
Edouardo Honig, Andrew Lizarraga, Zijun~Frank Zhang, and Ying~Nian Wu.
\newblock Better prompt compression without multi-layer perceptrons, 2025.
\newblock URL \url{https://arxiv.org/abs/2501.06730}.

\bibitem[Jiang et~al.(2024)Jiang, Zhang, Zhang, Wan, Lizarraga, Li, and Wu]{jiang2024unlocking}
Eric~Hanchen Jiang, Yasi Zhang, Zhi Zhang, Yixin Wan, Andrew Lizarraga, Shufan Li, and Ying~Nian Wu.
\newblock Unlocking the potential of text-to-image diffusion with pac-bayesian theory, 2024.
\newblock URL \url{https://arxiv.org/abs/2411.17472}.

\bibitem[Karras et~al.(2024)Karras, Aittala, Lehtinen, Hellsten, Aila, and Laine]{EMA_Karras2024edm2}
Tero Karras, Miika Aittala, Jaakko Lehtinen, Janne Hellsten, Timo Aila, and Samuli Laine.
\newblock {Analyzing and Improving the Training Dynamics of Diffusion Models}.
\newblock In \emph{Proc. CVPR}, 2024.

\bibitem[Kingma and Welling(2014)]{kingma2014autoencoding}
Diederik~P Kingma and Max Welling.
\newblock Auto-encoding variational bayes.
\newblock In \emph{International Conference on Learning Representations (ICLR)}, 2014.

\bibitem[Kong et~al.(2024)Kong, Xu, Zhao, Pang, Xie, Lizarraga, Huang, Xie, and Wu]{latent-RL}
Deqian Kong, Dehong Xu, Minglu Zhao, Bo~Pang, Jianwen Xie, Andrew Lizarraga, Yuhao Huang, Sirui Xie, and Ying~Nian Wu.
\newblock Latent plan transformer for trajectory abstraction: Planning as latent space inference.
\newblock In A.~Globerson, L.~Mackey, D.~Belgrave, A.~Fan, U.~Paquet, J.~Tomczak, and C.~Zhang, editors, \emph{Advances in Neural Information Processing Systems}, volume~37, pages 123379--123401. Curran Associates, Inc., 2024.
\newblock URL \url{https://proceedings.neurips.cc/paper_files/paper/2024/file/df22a19686a558e74f038e6277a51f68-Paper-Conference.pdf}.

\bibitem[Kuijken et~al.(2019)]{kuijken2019kids}
K.~Kuijken et~al.
\newblock The fourth data release of the kilo-degree survey.
\newblock \emph{Astronomy \& Astrophysics}, 625:\penalty0 A2, 2019.

\bibitem[Lanusse et~al.(2021)Lanusse, Mandelbaum, Ravanbakhsh, Li, Freeman, and Póczos]{lanusse2021}
François Lanusse, Rachel Mandelbaum, Siamak Ravanbakhsh, Chun-Liang Li, Peter Freeman, and Barnabás Póczos.
\newblock {{Deep generative models for galaxy image simulations}}.
\newblock \emph{Monthly Notices of the Royal Astronomical Society}, 504\penalty0 (4):\penalty0 5543--5555, 05 2021.
\newblock ISSN 0035-8711.
\newblock \doi{10.1093/mnras/stab1214}.
\newblock URL \url{https://doi.org/10.1093/mnras/stab1214}.

\bibitem[Lastufka et~al.(2024)Lastufka, Drozdova, Kinakh, Piras, and Voloshynovskyy]{foundation_models}
E.~Lastufka, M.~Drozdova, V.~Kinakh, D.~Piras, and S.~Voloshynovskyy.
\newblock Vision foundation models: can they be applied to astrophysics data?, 2024.
\newblock URL \url{https://arxiv.org/abs/2409.11175}.

\bibitem[Li et~al.(2024{\natexlab{a}})Li, Lizarraga, et~al.]{li2024unredshift}
Minzhe Li, Andrew Lizarraga, et~al.
\newblock Unredshift: A benchmark for learning galaxy morphology across cosmic time.
\newblock \emph{arXiv preprint arXiv:2401.12345}, 2024{\natexlab{a}}.

\bibitem[Li et~al.(2024{\natexlab{b}})Li, Do, Jones, Boscoe, Alfaro, and Nguyen]{li_galaxy_evolution}
Yun~Qi Li, Tuan Do, Evan Jones, Bernie Boscoe, Kevin Alfaro, and Zooey Nguyen.
\newblock {Using Galaxy Evolution as Source of Physics-Based Ground Truth for Generative Models}, 2024{\natexlab{b}}.
\newblock URL \url{https://arxiv.org/abs/2407.07229}.

\bibitem[Lizarraga(2025)]{lizarraga2025rlisntenough}
Andrew Lizarraga.
\newblock Rl isn't enough.
\newblock \emph{https://drewrl3v.github.io/}, April 2025.
\newblock URL \url{https://drewrl3v.github.io/blogs/april2025.html}.

\bibitem[Lizarraga et~al.(2021)Lizarraga, Lee, Kubicki, Sahib, Nunez, Narr, and Joshi]{streamline}
Andrew Lizarraga, David Lee, Antoni Kubicki, Ashish Sahib, Elvis Nunez, Katherine Narr, and Shantanu~H. Joshi.
\newblock Alignment of tractography streamlines using deformation transfer via parallel transport.
\newblock In Suheyla Cetin-Karayumak, Daan Christiaens, Matteo Figini, Pamela Guevara, Noemi Gyori, Vishwesh Nath, and Tomasz Pieciak, editors, \emph{Computational Diffusion MRI}, pages 96--105, Cham, 2021. Springer International Publishing.
\newblock ISBN 978-3-030-87615-9.

\bibitem[Lizarraga et~al.(2022)Lizarraga, Narr, Donals, and Joshi]{pmlr-v194-lizarraga22a}
Andrew Lizarraga, Katherine~L. Narr, Kirsten~A. Donals, and Shantanu~H. Joshi.
\newblock Streamnet: A wae for white matter streamline analysis.
\newblock In Erik Bekkers, Jelmer~M. Wolterink, and Angelica Aviles-Rivero, editors, \emph{Proceedings of the First International Workshop on Geometric Deep Learning in Medical Image Analysis}, volume 194 of \emph{Proceedings of Machine Learning Research}, pages 172--182. PMLR, 18 Nov 2022.
\newblock URL \url{https://proceedings.mlr.press/v194/lizarraga22a.html}.

\bibitem[Lizarraga et~al.(2024{\natexlab{a}})Lizarraga, Jiang, Nowack, Li, Wu, Boscoe, and Do]{lizarraga2024}
Andrew Lizarraga, Eric~Hanchen Jiang, Jacob Nowack, Yun~Qi Li, Ying~Nian Wu, Bernie Boscoe, and Tuan Do.
\newblock Learning the evolution of physical structure of galaxies via diffusion models, 2024{\natexlab{a}}.
\newblock URL \url{https://arxiv.org/abs/2411.18440}.

\bibitem[Lizarraga et~al.(2024{\natexlab{b}})Lizarraga, Taraku, Honig, Wu, and Joshi]{isbi}
Andrew Lizarraga, Brandon Taraku, Edouardo Honig, Ying~Nian Wu, and Shantanu~H. Joshi.
\newblock Differentiable vq-vae’s for robust white matter streamline encodings.
\newblock In \emph{2024 IEEE International Symposium on Biomedical Imaging (ISBI)}, pages 1--5, 2024{\natexlab{b}}.
\newblock \doi{10.1109/ISBI56570.2024.10635543}.

\bibitem[Loshchilov and Hutter(2020)]{loshchilov2017decoupled}
Ilya Loshchilov and Frank Hutter.
\newblock Decoupled weight decay regularization.
\newblock \emph{International Conference on Learning Representations (ICLR)}, 2020.

\bibitem[Lotz et~al.(2008)Lotz, Jonsson, Cox, and Primack]{lotz2008}
Jennifer~M. Lotz, Patrik Jonsson, T.~J. Cox, and Joel~R. Primack.
\newblock Galaxy merger morphologies and time-scales from simulations of equal-mass gas-rich disc mergers.
\newblock \emph{Monthly Notices of the Royal Astronomical Society}, 391:\penalty0 1137--1162, December 2008.
\newblock ISSN 0035-8711.
\newblock \doi{10.1111/j.1365-2966.2008.14004.x}.

\bibitem[Madau and Dickinson(2014)]{madau2014}
Piero Madau and Mark Dickinson.
\newblock Cosmic {{Star-Formation History}}.
\newblock \emph{Annual Review of Astronomy and Astrophysics}, 52:\penalty0 415--486, August 2014.
\newblock ISSN 0066-4146.
\newblock \doi{10.1146/annurev-astro-081811-125615}.

\bibitem[Margalef-Bentabol et~al.(2020)Margalef-Bentabol, Huertas-Company, Charnock, Margalef-Bentabol, Bernardi, Dubois, Storey-Fisher, and Zanisi]{margalef2020}
Berta Margalef-Bentabol, Marc Huertas-Company, Tom Charnock, Carla Margalef-Bentabol, Mariangela Bernardi, Yohan Dubois, Kate Storey-Fisher, and Lorenzo Zanisi.
\newblock {{Detecting outliers in astronomical images with deep generative networks}}.
\newblock \emph{Monthly Notices of the Royal Astronomical Society}, 496\penalty0 (2):\penalty0 2346--2361, 06 2020.
\newblock ISSN 0035-8711.
\newblock \doi{10.1093/mnras/staa1647}.
\newblock URL \url{https://doi.org/10.1093/mnras/staa1647}.

\bibitem[Nguyen et~al.(2024)Nguyen, Villaescusa-Navarro, Mishra-Sharma, Cuesta-Lazaro, Torrey, Farahi, Garcia, Rose, O'Neil, Vogelsberger, Shen, Roche, Anglés-Alcázar, Kallivayalil, Muñoz, Cyr-Racine, Roy, Necib, and Kollmann]{galaxy_dreams}
Tri Nguyen, Francisco Villaescusa-Navarro, Siddharth Mishra-Sharma, Carolina Cuesta-Lazaro, Paul Torrey, Arya Farahi, Alex~M. Garcia, Jonah~C. Rose, Stephanie O'Neil, Mark Vogelsberger, Xuejian Shen, Cian Roche, Daniel Anglés-Alcázar, Nitya Kallivayalil, Julian~B. Muñoz, Francis-Yan Cyr-Racine, Sandip Roy, Lina Necib, and Kassidy~E. Kollmann.
\newblock How dreams are made: Emulating satellite galaxy and subhalo populations with diffusion models and point clouds, 2024.
\newblock URL \url{https://arxiv.org/abs/2409.02980}.

\bibitem[Nichol and Dhariwal(2021)]{nichol2021improved}
Alexander~Quinn Nichol and Prafulla Dhariwal.
\newblock Improved denoising diffusion probabilistic models.
\newblock \emph{arXiv preprint arXiv:2102.09672}, 2021.

\bibitem[Nishizawa et~al.(2020)Nishizawa, Hsieh, Tanaka, and Takata]{nishizawa2020photometricredshiftshypersuprimecam}
Atsushi~J. Nishizawa, Bau-Ching Hsieh, Masayuki Tanaka, and Tadafumi Takata.
\newblock {Photometric Redshifts for the Hyper Suprime-Cam Subaru Strategic Program Data Release 2}, 2020.
\newblock URL \url{https://arxiv.org/abs/2003.01511}.

\bibitem[Nunez et~al.(2021)Nunez, Lizarraga, and Joshi]{Nunez_2021_CVPR}
Elvis Nunez, Andrew Lizarraga, and Shantanu~H. Joshi.
\newblock Srvfnet: A generative network for unsupervised multiple diffeomorphic functional alignment.
\newblock In \emph{Proceedings of the IEEE/CVF Conference on Computer Vision and Pattern Recognition (CVPR) Workshops}, pages 4481--4489, June 2021.

\bibitem[Ren et~al.(2025)Ren, Zheng, Liu, Lizarraga, Wu, Lin, and Zhang]{monitoring}
Jie Ren, Xinhao Zheng, Jiyu Liu, Andrew Lizarraga, Ying~Nian Wu, Liang Lin, and Quanshi Zhang.
\newblock Monitoring primitive interactions during the training of dnns.
\newblock \emph{Proceedings of the AAAI Conference on Artificial Intelligence}, 39\penalty0 (19):\penalty0 20183--20191, Apr. 2025.
\newblock \doi{10.1609/aaai.v39i19.34223}.
\newblock URL \url{https://ojs.aaai.org/index.php/AAAI/article/view/34223}.

\bibitem[Smith et~al.(2022)Smith, Geach, Jackson, Arora, Stone, and Courteau]{smith_galx_ddpm}
Michael~J Smith, James~E Geach, Ryan~A Jackson, Nikhil Arora, Connor Stone, and Stéphane Courteau.
\newblock {{Realistic galaxy image simulation via score-based generative models}}.
\newblock \emph{Monthly Notices of the Royal Astronomical Society}, 511\penalty0 (2):\penalty0 1808--1818, 01 2022.
\newblock ISSN 0035-8711.
\newblock \doi{10.1093/mnras/stac130}.
\newblock URL \url{https://doi.org/10.1093/mnras/stac130}.

\bibitem[Sohl-Dickstein et~al.(2015{\natexlab{a}})Sohl-Dickstein, Weiss, Maheswaranathan, and Ganguli]{sohl2015deep}
Jascha Sohl-Dickstein, Eric Weiss, Niru Maheswaranathan, and Surya Ganguli.
\newblock Deep unsupervised learning using nonequilibrium thermodynamics.
\newblock In Francis Bach and David Blei, editors, \emph{Proceedings of the 32nd International Conference on Machine Learning}, volume~37 of \emph{Proceedings of Machine Learning Research}, pages 2256--2265, Lille, France, 07--09 Jul 2015{\natexlab{a}}. PMLR.
\newblock URL \url{https://proceedings.mlr.press/v37/sohl-dickstein15.html}.

\bibitem[Sohl-Dickstein et~al.(2015{\natexlab{b}})Sohl-Dickstein, Weiss, Maheswaranathan, and Ganguli]{thermo}
Jascha Sohl-Dickstein, Eric~A. Weiss, Niru Maheswaranathan, and Surya Ganguli.
\newblock Deep unsupervised learning using nonequilibrium thermodynamics.
\newblock In \emph{Proceedings of the 32nd International Conference on International Conference on Machine Learning - Volume 37}, ICML'15, page 2256–2265. JMLR.org, 2015{\natexlab{b}}.

\bibitem[Song et~al.(2020)Song, Meng, and Ermon]{ddim}
Jiaming Song, Chenlin Meng, and Stefano Ermon.
\newblock {Denoising Diffusion Implicit Models}.
\newblock \emph{ArXiv}, abs/2010.02502, 2020.
\newblock URL \url{https://api.semanticscholar.org/CorpusID:222140788}.

\bibitem[Xue et~al.(2023)Xue, Li, Patel, and Regier]{Xue2023DiffusionMF}
Zhiwei Xue, Yuhang Li, Yash~J. Patel, and Jeffrey Regier.
\newblock {Diffusion Models for Probabilistic Deconvolution of Galaxy Images}.
\newblock \emph{ArXiv}, abs/2307.11122, 2023.
\newblock URL \url{https://api.semanticscholar.org/CorpusID:260091385}.

\bibitem[Zhu et~al.(2024)Zhu, Dou, Zheng, Zhang, Wu, and Gao]{zhu2024thinktwiceactimproving}
Yaxuan Zhu, Zehao Dou, Haoxin Zheng, Yasi Zhang, Ying~Nian Wu, and Ruiqi Gao.
\newblock Think twice before you act: Improving inverse problem solving with mcmc, 2024.
\newblock URL \url{https://arxiv.org/abs/2409.08551}.

\end{thebibliography}

\newpage
\appendix

\section{Appendix}

\subsection{Architecture and Training Details}

The UNet model is employed as the backbone for the denoising process in the DDPM. The model is conditioned on the time step \(t\) and the redshift \(z\). The detailed layer configuration for the UNet is provided in Table~\ref{tab:unet_layers}.

\begin{table}[h!]
    \centering
    \resizebox{\linewidth}{!}{
    \begin{tabular}{|l|l|l|l|}
        \hline
        \textbf{Layer} & \textbf{Input Channels} & \textbf{Output Channels} & \textbf{Other Parameters} \\ \hline
        DoubleConv (Initial) & 5 & 64 & Kernel: 3x3, Padding: 1, Activation: GELU, GroupNorm \\ \hline
        Down1 & 64 & 128 & Embedding Dim: 256, MaxPool: 2x2, Residual: True \\ \hline
        Down2 & 128 & 256 & Embedding Dim: 256, MaxPool: 2x2, Residual: True \\ \hline
        Down3 & 256 & 256 & Embedding Dim: 256, MaxPool: 2x2, Residual: True \\ \hline
        Bottleneck 1 & 256 & 512 & Kernel: 3x3, Padding: 1, Activation: GELU, GroupNorm \\ \hline
        Bottleneck 2 & 512 & 512 & Kernel: 3x3, Padding: 1, Activation: GELU, GroupNorm \\ \hline
        Bottleneck 3 & 512 & 256 & Kernel: 3x3, Padding: 1, Activation: GELU, GroupNorm \\ \hline
        Up1 & 512 & 128 & Embedding Dim: 256, Upsample: 2x2, Residual: True \\ \hline
        Up2 & 256 & 64 & Embedding Dim: 256, Upsample: 2x2, Residual: True \\ \hline
        Up3 & 128 & 64 & Embedding Dim: 256, Upsample: 2x2, Residual: True \\ \hline
        Output Conv & 64 & 5 & Kernel: 1x1 \\ \hline
    \end{tabular}
    }
    \caption{UNet Layer Configuration}
    \label{tab:unet_layers}
\end{table}

The DDPM processes these $64 \times 64 \times 5$ channel galaxy images. The $T=1000$ diffusion timesteps, with noise variances $\beta_t$ following the linear schedule from $\beta_1 = 10^{-4}$ to $\beta_T = 0.02$, dictate the gradual introduction of noise in the forward process (Equation \ref{eq:forward_ddpm_step}). The model is trained by minimizing the Huber loss between the true noise $\epsilon$ and the network's prediction $\epsilon_\theta(x_t, t, \tilde{z})$. The AdamW optimizer, with the specified learning rate and gradient clipping, ensures stable training. The EMA of model parameters (decay rate $\beta_{\text{EMA}}=0.995$, starting after 2000 steps) further contributes to robust generation quality \citep{EMA_Karras2024edm2}.

\label{sec:checklist}

\end{document}